\documentclass[a4paper,11pt]{article}
\usepackage{pos, lineno}
\title{Femtoscopy using L\'evy-distributed sources at NA61/SHINE}

\author*[a, b]{Barnab\'as P\'orfy}
\onbehalf{for the NA61/SHINE Collaboration}

\affiliation[a]{E\"otv\"os Lor\'and University,\\
  Budapest, Hungary}

\affiliation[b]{HUN-REN Wigner Research Centre for Physics\\
Budapest, Hungary}

\emailAdd{barnabas.porfy@cern.ch}
\emailAdd{porfy.barnabas@wigner.hun-ren.hu}

\abstract{In the recent years, research studies in high-energy physics have confirmed the creation of the strongly interacting quark-gluon plasma (sQGP) in ultra-relativistic nucleus-nucleus collisions. NA61/SHINE at CERN SPS investigates hadronic matter properties by varying collision energy ($\sqrt{s_{\rm{NN}}} \approx 5.3, 6.2, 7.7, 8.8, 12$, and 16.8 GeV) and systems (such as p+p, p+Pb, Be+Be, Ar+Sc, Xe+La, Pb+Pb). Utilizing femtoscopic correlations, we can unveil the space-time structure of the hadron emitting source. Our focus is on femtoscopic correlations in small to intermediate systems, comparing measurements with source calculations based on Lévy-distributed sources, to explore the pair transverse mass dependence of the Lévy source parameters. The Lévy exponent $\alpha$ is of particular significance, which characterizes the shape of the source and may be connected to the critical exponent $\eta$ near the critical point. Our analysis will reveal that the L\'evy shape parameter, $\alpha$, has a slight non-monotonic behaviour as a function of collision energy and that we see a deviation from Gaussian sources. Finally, it will be shown that there is no indication of the critical point at any of the investigated energies.}

\FullConference{42nd International Conference on High Energy Physics (ICHEP2024)\\
18-24 July 2024\\
Prague, Czech Republic\\}

\begin{document}
\maketitle

\section{Introduction}
The NA61/SHINE experiment, located on the CERN SPS H2 beam line, is a fixed target experiment, using multiple Time Projection Chambers (TPCs) for large-acceptance hadron spectroscopy~\cite{Abgrall:2014xwa}. Its detector setup allows tracking down to $p_{\rm{T}} \approx 0$ GeV/\textit{c}. NA61/SHINE primarily aims to explore the phase diagram of strongly interacting matter across various temperatures and baryon-chemical potentials through system and beam energy scans. This work focuses on $^{40}$Ar+$^{45}$Sc collisions at the three lowest available energies (13, 19, and 30\textit{A} GeV/\textit{c})
at 0--10\% centrality, and comparing with previous results~\cite{NA61SHINE:2023qzr,Porfy:2023yii, Porfy:2024kbk} using femtoscopy correlations with spherically symmetric L\'evy distributions: 
\begin{equation}\label{eq:levydistr}
\mathcal{L}(\alpha,R,\boldsymbol{r})=\frac{1}{(2\pi)^3} \int \rm{d}^3 \boldsymbol{ \zeta} e^{i \boldsymbol{ \zeta} \boldsymbol{r}} e^{-\frac{1}{2}| \boldsymbol{ \zeta} R|^{\alpha}},
\end{equation}
where $R$ is the L\'evy scale parameter, $\alpha$ is the L\'evy stability index, $\boldsymbol{r}$ is the vector of spatial coordinates. Bose-Einstein momentum correlations are related to the particle source function $S(x)$ via the equation $C(q) \cong 1 + | \tilde{S}(q) |^2$, see Ref.~\cite{NA61SHINE:2023qzr} for details. The L\'evy distribution generalizes Gaussian distributions ($\alpha = 2$) and captures effects like critical fluctuations~\cite{Csorgo:2008ayr}. For $\alpha < 2$, the L\'evy distribution exhibits a power-law tail $\sim r^{-(1 + \alpha)}$. Critical behavior is related to the exponent $\eta$, which also follows a power-law tail and has been linked to the 3D Ising model, where $\eta = 0.03631 \pm 0.00003$~\cite{El-Showk:2014dwa} and to the 3D Ising model with random external field, resulting in $\eta = 0.50 \pm 0.05$~\cite{Rieger:1995aa}. Other phenomena could have influences on the source shape, such as QCD jets, anomalous diffusion, and others, as discussed in Refs.~\cite{Metzler:1999zz,Csorgo:2003uv, Csorgo:2004sr, Kincses:2022eqq, Korodi:2022ohn, Kincses:2024lnv}. For overview of recent results, see Ref.~\cite{Csanad:2024hva}.

The L\'evy exponent can be measured utilizing the following femtoscopic correlation function to data: 
\begin{equation}\label{eq:corrfunc}
C^0_2(q) = 1 + \lambda \cdot e^{-(qR)^\alpha},
\end{equation}
where $C^0_2(q)$ is the correlation function in absence of interaction and $\lambda$ is the correlation strength, often interpreted through the core-halo model~\cite{Csorgo:1995bi, Csorgo:1999sj} and is related to the core and halo pion multiplicities (denoted by \textit{N}): 
\begin{equation}\label{eq:lambda}
\lambda = \left(N_{\rm{core}}/N_{\textnormal{total}}\right)^2.
\end{equation}

We investigate the combination of positive ($\pi^+\pi^+$) and negative ($\pi^-\pi^-$) pion pairs, identified via energy loss in the TPCs (d$E$/d$x$) and comparison with the Bethe-Bloch curves. Track merging was eliminated with momentum-based two-track distance cuts~\cite{Czopowicz:2022nzy}. Centrality was selected via forward energy measurements. Pion pairs were grouped into four to eight (energy dependent) $K_{\rm{T}}$ bins (0-500 MeV/c). We address Coulomb repulsion effects on like-charged pairs using the correction factor $K_{\rm{Coulomb}}(q)$:
\begin{equation}\label{eq:coulombdef}
K_{\rm{Coulomb}}(q) = \frac{C^{\rm{Coulomb}}_2(q)}{C^0_2(q)},
\end{equation}
where $C^{\rm{Coulomb}}_2(q)$ is calculated numerically~\cite{Kincses:2019rug, Csanad:2019lkp}. We are utilizing a novel method in our analysis for estimating the effect of Coulomb repulsion, presented in Ref.~\cite{Nagy:2023zbg}. The correlation function, corrected for Coulomb interaction, using the Bowler-Sinyukov method~\cite{Sinyukov:1998fc}, takes the form:
\begin{equation}\label{eq:fitfunc}
C_2(q) = N\cdot \left( 1 + \epsilon \cdot q\right) \cdot \left[1 - \lambda + \lambda \cdot \left(1 + e^{-|qR|^\alpha} \right) \cdot K_{\rm{Coulomb}}(q) \right],
\end{equation}
where $N$ is introduced as normalization parameter and $\varepsilon$ is introduced to describe the residual background in the form of $\left(1 + \varepsilon \cdot q \right)$. The Coulomb correction is applied in the pair-center-of-mass system (PCMS), with corrections for the longitudinally co-moving system (LCMS)~\cite{Kurgyis:2020vbz}.


\section{Results}
\begin{figure}
\centerline{\includegraphics[width=.7\textwidth]{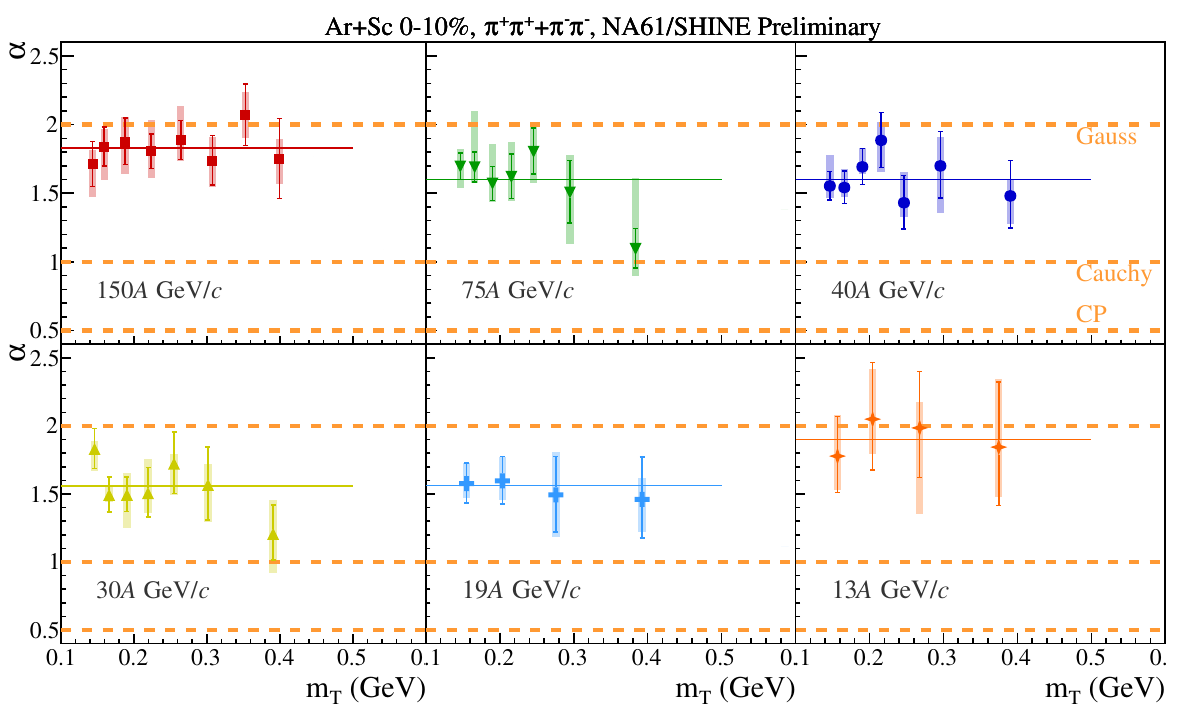}}
\caption{The fit parameters, for $0$--$10$\% central Ar+Sc at all available energies, as a function of $m_{\rm{T}}$. Special cases corresponding to a Gaussian ($\alpha=2$) or a Cauchy ($\alpha=1$) source are shown, as well as $\alpha=0.5$, the conjectured value corresponding to the critical endpoint, while the constant $\alpha$ fit is shown with a solid line. Boxes denote systematic uncertainties, bars represent statistical uncertainties.\label{fig:resultsA}}
\end{figure}

The physical parameters $\alpha$, $R$, and $\lambda$ were measured in bins of pair transverse momentum, $K_{\rm{T}}$, at 13, 19, and 30\textit{A} GeV/$c$ by fitting the measured correlation functions using Eq{.}(\ref{eq:fitfunc}). Their dependence on transverse mass $m_{\rm{T}} = \sqrt{m^2c^4+K_{\rm{T}}^2c^2}$ was studied and compared to previous NA61/SHINE results, taken from Refs.~\cite{NA61SHINE:2023qzr, Porfy:2023yii, Porfy:2024kbk}. The stability exponent $\alpha$ indicates the strength of the source tail, with results ranging from 1.5 to 2.0, with the lowest energy being comparable to 150\textit{A} GeV/$c$, implying sources that are closer to Gaussian (Fig.~\ref{fig:resultsA}). These values are higher than the conjectured value at the critical endpoint ($\alpha \approx 0.5$), with a non-monotonic trend emerging around $\sqrt{s_{\rm{NN}}} \approx 6-8$ GeV. It is interesting to compare our $\alpha$ values to results from other experiments, see details in Ref.~\cite{Csanad:2024hva}.

The L\'evy scale parameter $R$ is related to the homogeneity scale of the pion source. A slight decrease in $R$ with $m_{\rm{T}}$ was observed, compatible with $R \sim 1/\sqrt{m_{\rm{T}}}$, consistently with hydrodynamical predictions ~\cite{Sinyukov:1994vg,Sinyukov1995} for Gaussian sources, based on transverse flow. It is particularly interesting that this results holds for L\'evy sources as well. A similar behavior was observed at RHIC and LHC energies~\cite{PHENIX:2017ino,Kincses:2024sin} and in simulations as well~\cite{Kincses:2022eqq}.

\begin{figure}
\centerline{\includegraphics[width=.7\textwidth]{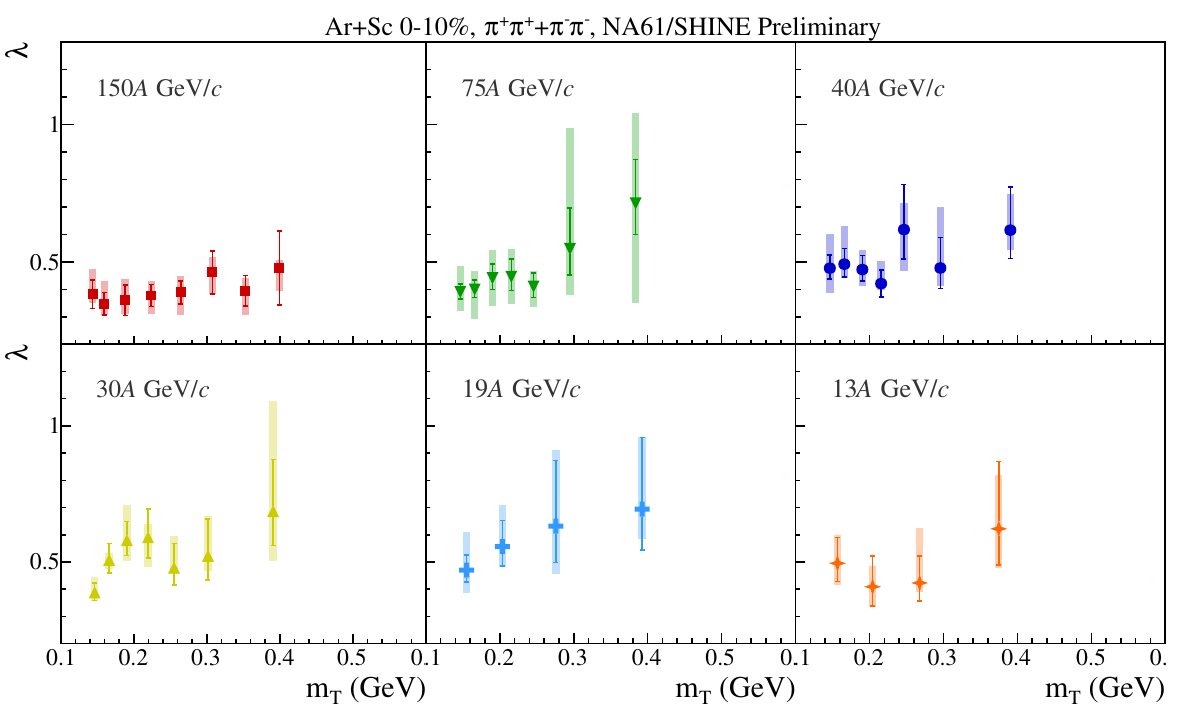}}
\caption{The fit parameters, for $0$--$10$\% central Ar+Sc at all available energies, as a function of $m_{\rm{T}}$. Boxes denote systematic uncertainties, bars represent statistical uncertainties.\label{fig:resultsL}}
\end{figure}

The correlation strength $\lambda$ shows a moderate $m_{\rm{T}}$ dependence but remains mostly constant, as shown in Fig.~\ref{fig:resultsL}. Our $\lambda$ values are lower than unity, suggesting a significant fraction of pions are decay products of long-lived resonances. At RHIC, a decrease of $\lambda$ at low $m_{\rm{T}}$ was seen and attributed to in-medium mass modification of $\eta'$~\cite{PHENIX:2017ino,PHENIX:2024vjp}. We do not see this effect in our results.

\begin{figure}
\centerline{\includegraphics[width=.6\textwidth]{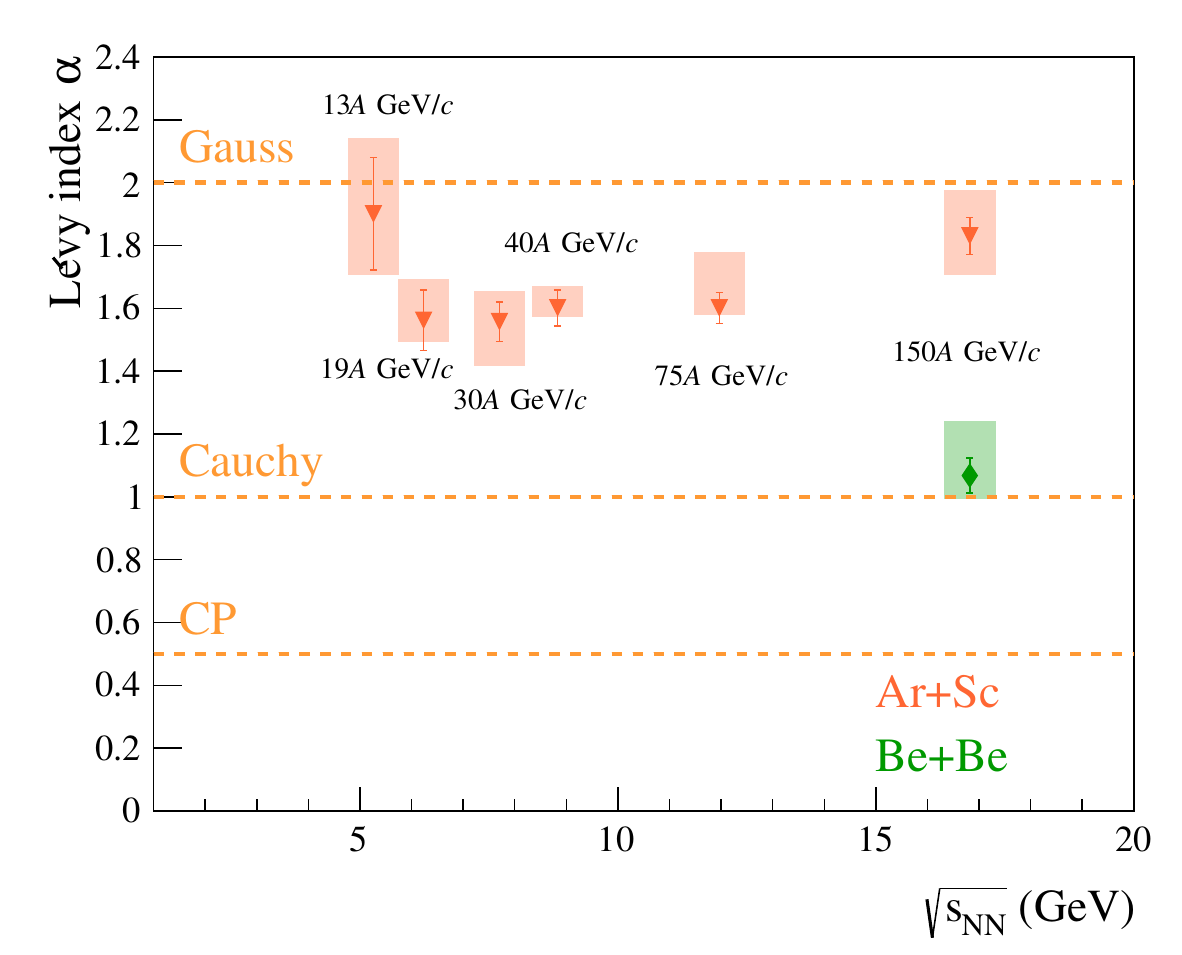}}
\caption{The constant fit to $\alpha$, for $0$--$20$ \% central Be+Be at 150\textit{A} GeV/\textit{c} and $0$--$10$\% central Ar+Sc at all available energies, as a function of $\sqrt{s_{\rm{NN}}}$. Special cases corresponding to a Gaussian ($\alpha=2$) or a Cauchy ($\alpha=1$) source are shown, as well as $\alpha=0.5$, the conjectured value corresponding to the critical endpoint. Boxes denote systematic uncertainties, bars represent statistical uncertainties.\label{fig:resultsAfit}}
\end{figure}

\begin{figure}
\centerline{\includegraphics[width=.7\textwidth]{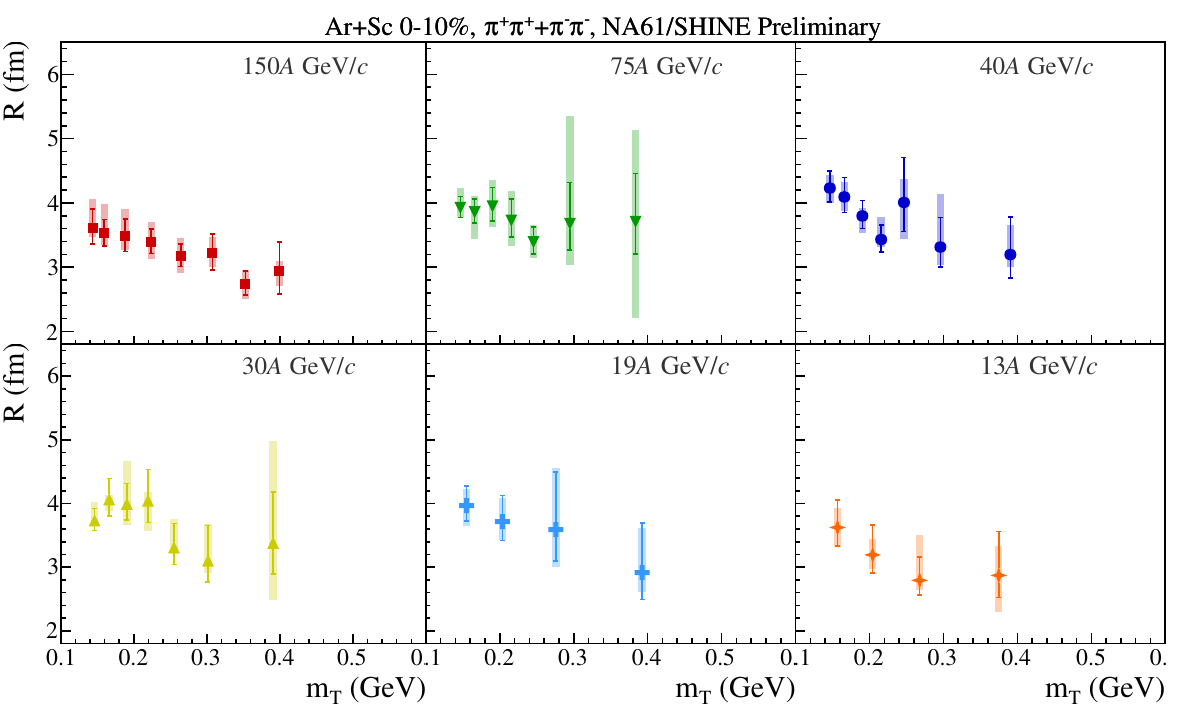}}
\caption{The fit parameters, for $0$--$10$\% central Ar+Sc at all available energies, as a function of $m_{\rm{T}}$. Boxes denote systematic uncertainties, bars represent statistical uncertainties.\label{fig:resultsR}}
\end{figure}

\section{Conclusion}
In this proceedings, we presented the NA61/SHINE measurements of one-dimensional two-pion femtoscopic correlation functions in Ar+Sc collisions at 13, 19, and 30\textit{A} GeV/\textit{c} beam momenta in 0\--10\% centrality collisions. We fitted these correlation functions based on L\'evy-shaped source distributions, and investigated the transverse mass dependence of the source parameters. We furthermore compared them to previous NA61/SHINE results. Additionally, the L\'evy scale parameter, $R$, exhibits a noticeable decrease with $m_{\rm{T}}$. Finally, the correlation strength parameter, $\lambda$, shows no significant dependence on $m_{\rm{T}}$, unlike RHIC results but consistent with earlier SPS measurements. In future studies, we plan to measure Bose-Einstein correlations in larger systems and at lower energies to further investigate the phase diagram of strongly interacting matter.

\section*{Acknowledgments}

The author acknowledges support of the DKOP-23 Doctoral Excellence Program of the Ministry for Culture and Innovation, and was furthermore supported by K-138136 and K-138152 grants of the National Research, Development and Innovation Fund. 
\bibliographystyle{na61Utphys}
\bibliography{pos}

\providecommand{\href}[2]{#2}\begingroup\raggedright\begin{thebibliography}{10}

\bibitem{Abgrall:2014xwa}
N.~Abgrall {\em et~al.}, {[NA61/SHINE} Collab.] \href{http://dx.doi.org/10.1088/1748-0221/9/06/P06005}{{\em JINST} {\bfseries 9} (2014) P06005},
\href{http://arxiv.org/abs/1401.4699}{{\ttfamily arXiv:1401.4699 [physics.ins-det]}}.

\bibitem{NA61SHINE:2023qzr}
H.~Adhikary {\em et~al.}, {[NA61/SHINE} Collab.] \href{http://dx.doi.org/10.1140/epjc/s10052-023-11997-8}{{\em Eur. Phys. J. C} {\bfseries 83} no.~10, (2023) 919}, \href{http://arxiv.org/abs/2302.04593}{{\ttfamily arXiv:2302.04593 [nucl-ex]}}.

\bibitem{Porfy:2023yii}
B.~Porfy, {[NA61/SHINE} Collab.] \href{http://dx.doi.org/10.3390/universe9070298}{{\em Universe} {\bfseries 9} no.~7, (2023) 298}, \href{http://arxiv.org/abs/2306.08696}{{\ttfamily arXiv:2306.08696 [nucl-ex]}}.

\bibitem{Porfy:2024kbk}
B.~Porfy, {[NA61/SHINE} Collab.], ``{Femtoscopy at NA61/SHINE using symmetric L\'evy sources in central $^{40}$Ar+$^{45}$Sc from 40$A$ GeV/$c$ to 150$A$ GeV/$c$},'' in {\em {23rd Zimanyi School Winter Workshop}}.
\newblock 6, 2024.
\newblock \href{http://arxiv.org/abs/2406.02242}{{\ttfamily arXiv:2406.02242 [nucl-ex]}}.

\bibitem{Csorgo:2008ayr}
T.~Cs{\"o}rg\H{o} \href{http://dx.doi.org/10.22323/1.076.0027}{{\em PoS} {\bfseries HIGH-PTLHC08} (2008) 027}, \href{http://arxiv.org/abs/0903.0669}{{\ttfamily arXiv:0903.0669 [nucl-th]}}.

\bibitem{El-Showk:2014dwa}
S.~El-Showk {\em et~al.} \href{http://dx.doi.org/10.1007/s10955-014-1042-7}{{\em J. Stat. Phys.} {\bfseries 157} (2014) 869}, \href{http://arxiv.org/abs/1403.4545}{{\ttfamily arXiv:1403.4545 [hep-th]}}.

\bibitem{Rieger:1995aa}
H.~Rieger \href{http://dx.doi.org/10.1103/PhysRevB.52.6659}{{\em Physical Review B} {\bfseries 52} no.~9, (1995) 6659}. \url{http://link.aps.org/doi/10.1103/PhysRevB.52.6659}.

\bibitem{Metzler:1999zz}
R.~Metzler, E.~Barkai, and J.~Klafter \href{http://dx.doi.org/10.1103/PhysRevLett.82.3563}{{\em Phys. Rev. Lett.} {\bfseries 82} (1999) 3563--3567}.

\bibitem{Csorgo:2003uv}
T.~Cs{\"o}rg\H{o}, S.~Hegyi, and W.~A. Zajc \href{http://dx.doi.org/10.1140/epjc/s2004-01870-9}{{\em Eur. Phys. J. C} {\bfseries 36} (2004) 67--78}, \href{http://arxiv.org/abs/nucl-th/0310042}{{\ttfamily arXiv:nucl-th/0310042}}.

\bibitem{Csorgo:2004sr}
T.~Cs{\"o}rg\H{o}, S.~Hegyi, T.~Novak, and W.~A. Zajc {\em Acta Phys. Polon. B} {\bfseries 36} (2005) 329--337.

\bibitem{Kincses:2022eqq}
D.~Kincses, M.~Stefaniak, and M.~Csan\'ad \href{http://dx.doi.org/10.3390/e24030308}{{\em Entropy} {\bfseries 24} no.~3, (2022) 308}, \href{http://arxiv.org/abs/2201.07962}{{\ttfamily arXiv:2201.07962 [hep-ph]}}.

\bibitem{Korodi:2022ohn}
B.~K\'orodi, D.~Kincses, and M.~Csan\'ad \href{http://dx.doi.org/10.1016/j.physletb.2023.138295}{{\em Phys. Lett. B} {\bfseries 847} (2023) 138295}, \href{http://arxiv.org/abs/2212.02980}{{\ttfamily arXiv:2212.02980 [nucl-th]}}.

\bibitem{Kincses:2024lnv}
D.~Kincses, M.~Nagy, and M.~Csan\'ad \href{http://arxiv.org/abs/2409.10373}{{\ttfamily arXiv:2409.10373 [nucl-th]}}.

\bibitem{Csanad:2024hva}
M.~Csan\'ad and D.~Kincses \href{http://dx.doi.org/10.3390/universe10020054}{{\em Universe} {\bfseries 10} no.~2, (2024) 54}, \href{http://arxiv.org/abs/2401.01249}{{\ttfamily arXiv:2401.01249 [hep-ph]}}.

\bibitem{Csorgo:1995bi}
T.~Cs{\"o}rg\H{o} and B.~L{\"o}rstad \href{http://dx.doi.org/10.1103/PhysRevC.54.1390}{{\em Phys. Rev. C} {\bfseries 54} (1996) 1390--1403}, \href{http://arxiv.org/abs/hep-ph/9509213}{{\ttfamily arXiv:hep-ph/9509213}}.

\bibitem{Csorgo:1999sj}
T.~Cs{\"o}rg\H{o} \href{http://dx.doi.org/10.1556/APH.15.2002.1-2.1}{{\em Acta Phys. Hung. A} {\bfseries 15} (2002) 1--80}, \href{http://arxiv.org/abs/hep-ph/0001233}{{\ttfamily arXiv:hep-ph/0001233}}.

\bibitem{Czopowicz:2022nzy}
T.~Czopowicz, {[NA61/SHINE} Collab.] \href{http://dx.doi.org/10.22323/1.400.0039}{{\em PoS} {\bfseries CPOD2021} (2022) 039}.

\bibitem{Kincses:2019rug}
D.~Kincses, M.~I. Nagy, and M.~Csan\'ad \href{http://dx.doi.org/10.1103/PhysRevC.102.064912}{{\em Phys. Rev. C} {\bfseries 102} no.~6, (2020) 064912}, \href{http://arxiv.org/abs/1912.01381}{{\ttfamily arXiv:1912.01381 [hep-ph]}}.

\bibitem{Csanad:2019lkp}
M.~Csan\'ad, S.~L\"ok\"os, and M.~Nagy \href{http://dx.doi.org/10.1134/S1063779620030089}{{\em Phys. Part. Nucl.} {\bfseries 51} no.~3, (2020) 238--242}, \href{http://arxiv.org/abs/1910.02231}{{\ttfamily arXiv:1910.02231 [hep-ph]}}.

\bibitem{Nagy:2023zbg}
M.~Nagy, A.~Purzsa, M.~Csan\'ad, and D.~Kincses \href{http://dx.doi.org/10.1140/epjc/s10052-023-12161-y}{{\em Eur. Phys. J. C} {\bfseries 83} no.~11, (2023) 1015}, \href{http://arxiv.org/abs/2308.10745}{{\ttfamily arXiv:2308.10745 [nucl-th]}}.

\bibitem{Sinyukov:1998fc}
Y.~Sinyukov {\em et~al.} \href{http://dx.doi.org/10.1016/S0370-2693(98)00653-4}{{\em Phys. Lett. B} {\bfseries 432} (1998) 248--257}.

\bibitem{Kurgyis:2020vbz}
B.~Kurgyis, D.~Kincses, M.~Nagy, and M.~Csan\'ad \href{http://dx.doi.org/10.3390/universe9070328}{{\em Universe} {\bfseries 9} no.~7, (2023) 328}, \href{http://arxiv.org/abs/2007.10173}{{\ttfamily arXiv:2007.10173 [nucl-th]}}.

\bibitem{Sinyukov:1994vg}
Y.~M. Sinyukov \href{http://dx.doi.org/10.1016/0375-9474(94)90700-5}{{\em Nucl. Phys. A} {\bfseries 566} (1994) 589C--592C}.

\bibitem{Sinyukov1995}
Y.~M. Sinyukov {\em Hot Hadronic Matter: Theory and Experiment} (1995) 309--322.

\bibitem{PHENIX:2017ino}
A.~Adare {\em et~al.}, {[PHENIX} Collab.] \href{http://dx.doi.org/10.1103/PhysRevC.97.064911}{{\em Phys. Rev. C} {\bfseries 97} no.~6, (2018) 064911}, \href{http://arxiv.org/abs/1709.05649}{{\ttfamily arXiv:1709.05649 [nucl-ex]}}.

\bibitem{Kincses:2024sin}
D.~Kincses, {[STAR} Collab.] \href{http://dx.doi.org/10.3390/universe10030102}{{\em Universe} {\bfseries 10} no.~3, (2024) 102}, \href{http://arxiv.org/abs/2401.11169}{{\ttfamily arXiv:2401.11169 [nucl-ex]}}.

\bibitem{PHENIX:2024vjp}
N.~J. Abdulameer {\em et~al.}, {[PHENIX} Collab.] \href{http://arxiv.org/abs/2407.08586}{{\ttfamily arXiv:2407.08586 [nucl-ex]}}.

\end{thebibliography}\endgroup

\end{document}